\documentclass[a4paper,usenatbib,useAMS]{mnras}

\usepackage[T1]{fontenc}
\usepackage{ae,aecompl}

\usepackage{graphicx}	
\usepackage{amsmath}	
\usepackage{amssymb}	
\usepackage{bm}		
\usepackage{pdflscape} 	
\usepackage{lscape}

\usepackage[justification=centering]{caption}
\usepackage{longtable}
\usepackage{footnote}
\usepackage{stfloats}
\usepackage{natbib}
\usepackage{subfigure}
\usepackage{float}
\usepackage{times}
\usepackage{txfonts}
\usepackage{color}
\usepackage{makecell}
\usepackage{caption}

\usepackage{etoolbox}
\usepackage{subfigure}
\usepackage{epsf}
\usepackage{longtable}


\title{The Event Rate and Luminosity Function of Fermi/GBM Gamma-Ray Bursts}

\graphicspath{{./}{figures/}}

\author[Liu et al.]{
	Y. Liu$^{1}$,
	Z. B. Zhang$^{1}$\thanks{zbzhang@qfnu.edu.cn}, X. F. Dong$^{2}$, L. B. Li$^{3,4,5}$\thanks{lilongbiao@hebeu.edu.cn}
	and X. Y. Du$^{1}$
	\\
	$^{1}$College of Physics and Engineering, Qufu Normal University, Qufu 273165, P. R. China\\
	$^{2}$School of Astronomy and Space Science, Nanjing University, Nanjing 210093, China\\
	$^{3}$School of Mathematics and Physics, Hebei University of Engineering, Handan 056005, P. R. China\\
	$^{4}$Hebei Computational Optical Imaging and Photoelectric Detection Technology Innovation Center, Hebei University of Engineering, Handan 056005, P. R. China\\
	$^{5}$Hebei International Joint Research Center for Computational Optical Imaging and Intelligent Sensing, Hebei University of Engineering, Handan 056005, P. R. China
}
\date{Accepted XXX. Received YYY; in original form ZZZ}

\pubyear{2025}

\begin{document}

\label{firstpage}
\pagerange{\pageref{firstpage}--\pageref{lastpage}}
\maketitle

\begin{abstract}
Luminosity function and event rate of Gamma-Ray Bursts (GRBs) are easily biased by the instrument and selection effects. We select 115 Fermi/GBM GRBs with good spectra fitted by a smoothly broken power-law function. The $\tau$-statistic method is used to describe how the luminosity evolves with redshift. The non-parametric Lynden-Bell's c$^{-}$ method has been applied to get the cumulative luminosity function and event rate which is compared with the star formation history. How the selection and instrument effects bias the deduced event rate has been carefully studied. We find that the event rate always exceeds the star formation rate (SFR) at lower redshift and matches with each other at higher redshift, which is independent of energy bands and consistent with previous findings of other satellites. Furthermore, it is found that sample completeness does not affect the deduced event rate too much as mentioned for the Swift lGRBs in Dong et al.. A triple power-law function has been used to fit the cumulative flux distribution and categorize the total sample into three subsamples of bright, medium and faint GRBs. We find that the event rates of bright GRBs, unlike medium and faint ones, comply with the SFR ideally, which indicates that these bright GRBs with higher luminosity are possibly produced from the core-collapse of massive stars.
\end{abstract}


\section{Introduction}

\label{sec:intro}

Gamma-ray bursts (GRBs) are the most violent gamma-ray explosions in the Univese, with super-high energies and short time-scale \citep{2020ApJ...902...40Z}.
Typically, according to the prompt emission duration $T_{90}$, GRBs can be classified into long GRBs (lGRBs, $T_{90} > 2\rm s$) and short GRBs (sGRBs, $T_{90}<2\rm s$) of CGRO/BATSE \citep{1993ApJ...413L.101K} and Swift/BAT \citep{2008A&A...484..293Z} detectors. It is worth mentioning that different observation instruments would influence the measurement of $T_{90}$, which leads  the $T_{90} $ boundary between lGRBs and sGRBs to be about 1 s for Swift/BAT \citep{2020ApJ...902...40Z, 2022ApJ...940....5D} and 1.27s for Fermi/GBM \citep{2014ApJS..211...12G}. The lGRBs have isotropic $\gamma$-ray energiea in the range of $10^{51}\sim10^{54} \rm erg$ and are usually thought to originate from the collapse of massive stars \citep{1993ApJ...405..273W, 1997ApJ...486L..71T, 1998ApJ...494L..45P, 1998MNRAS.294L..13W, 2000AIPC..526..658L, 2001ApJ...548..522P, 2004IJMPA..19.2385Z, 2006ARA&A..44..507W}, which allows their event rates to follow the SFR in theory \citep{2010ApJ...711..495B, 2015ApJS..218...13Y, 2016A&A...587A..40P, 2022MNRAS.513.1078D, 2024ApJ...967L..30D}. SGRBs are gerenally believed to originate from merger of compact binary systems \citep{1992ApJ...395L..83N, 2013PhRvD..88d1503K, 2023ApJ...955...98L}, which has been verified by the detection of gravitational wave (GW) events GW170817 and sGRB 170817A \citep{2017ApJ...848L..13A, 2020MNRAS.499L..96P, 2022ApJ...932L...7C, 2023ApJ...952..157A, 2023A&A...669A..36V}. The relationships between the event rate of sGRBs and the delayed star formation models have been investigated \citep{2013A&A...558A..22H, 2020ApJ...902L..42W, 2022MNRAS.516.1654L,2021MNRAS.501..157Z,2025ApJ...987L..13D}. Meanwhile, other possible physical models including the binary-driven hypernova model \citep{2012ApJ...758L...7R, 2013IJMPD..2260009R, 2015ApJ...812..100B} have also been proposed.
With the accumulation of observations, the emerging observational and physical pictures, for example the ambiguous time profiles observed in both lGRBs and sGRBs, are becoming more and more complex for some bursts \citep{2021NatAs...5..911Z, 2022ApJ...932....1R, 2023NatAs...7..976L, 2024Natur.626..737L}. In particular, GRB 221009A as an extreme high-energy \citep{2023SciA....9J2778C,2025JHEAp..45..392Z} and nearby event has very large isotropic energy up to $10^{55}$ erg \citep{2023ApJ...952L..42L} that may be attributed to a narrow structured jet core \citep{2023Sci...380.1390L} or more extreme nenegy reservoir of central engine \citep{2023ApJ...949L...4L}. Although GRB 221009A possesses the peculiar observational features, it is still an ordinary burst compared with those normal lGRBs in many aspects, indicating their common progenitor of core-collapse \citep[e.g.][]{2023ApJ...949L...4L}.

The interstellar environment of kilonovae or supernovae associated with GRBs would influence both the GRB event rate and the nearby star formation rate \citep{2023ApJ...954L..17Y, 2024ApJ...961..201L}. Several other effects, such as neutral hydrogen column densities \citep{2019ApJ...885...47R}, jet structure  \citep{2024ApJ...967L...4H} and occultation by the Earth (data unavailability) \citep{2024arXiv240103589D} have been also considered as possible factors. In the observations, there are several selection effects can significantly influence the event rate and the redshifit $z$ detection rate of GRBs, such as Malmquist bias and the observation threshold of the detectors \citep{2007NewAR..51..539C}. In addition, the classification of low/high-luminosity GRBs \citep{2022MNRAS.513.1078D} was found to play an important role on the low-redshift excess of GRB event rate with regard to the SFR.

Several methods have been proposed to estimate the GRB event rates of diverse samples and compare them with the SFR. The $log N - log P$ distribution as a parametric method that has been used to evaluate the relationship between GRB event rate and the SFR and found to be essentially a convolution of luminosity and redshift \citep{2010ApJ...711..495B, 2018MNRAS.477.4275P, 2021ApJ...917...24Z,  2024MNRAS.528.5309K}. Lynden-Bell's c$^{-}$ method \citep{1971MNRAS.155...95L} is considered to derive the luminosity function $\psi (L)$ and the event rate $\rho (z)$ of GRB samples with known redshift and luminosity without any assumptions. Furthermore, this method can break the degeneracy between $\psi (L)$ and $\rho (z)$, so that both of them can be accurately determined. Note that Lynden-Bell's c$^{-}$ method has been applied applied in some relevant astronomy fields, such as lGRBs \citep{2016A&A...587A..40P, 2018ApJ...852....1Z, 2021RAA....21..254L, 2022MNRAS.513.1078D, 2023ApJ...958...37D, 2024MNRAS.527.7111L}, Fast Radio Bursts (FRBs, \citealt{2019JHEAp..23....1D,2025A&A...698A..18Z})  and Active Galactic Nuclei (AGNs, \citealt{2011ApJ...743..104S, 2021ApJ...913..120Z}, Rong, D. H. et al. 2025). Regarding the lGRB rate, some authors claimed that it should exceed the SFR at lower redshift \citep{2015ApJS..218...13Y,2015ApJ...806...44P,2017ApJ...837...17L,2019MNRAS.488.5823L,2020MNRAS.493.1479L,2020MNRAS.498.5041L,2021ApJ...908...83T,2024IJAA...14...20H}, while other authors expected the derived lGRB rate to match the SFR within the whole redshift range \citep[e.g.][]{2016A&A...587A..40P,2016ApJ...817....7P}.  In this work, we will continue to adopt the non-parametric method to investigate the connection between the GRB rate and the SFR in order to check the dependence of the GRB rates on the energy bands of detectors and how the GRB rates are affected by different factors.


The paper is organized as follows. We introduce the sample selection criteria  in Section \ref{sec:two} and the non-parametric method for the estimate of GRB event rate in Section \ref{method}. Our results are presented in Section \ref{sec:three}. A brief summary is presented in Section \ref{sec:four}.

\section{Sample selection} \label{sec:two}
In contrast with Swift/BAT detector (15-150 keV), Fermi/GBM has a wider energy range of 8 keV-40 MeV, a larger field of view and near real-time position capability \citep{2014ApJS..211...12G}, which enables it to detect more GRBs than before. Here, we collect 115 Fermi/GBM GRBs with peak energy flux and redshift in the HEASARC Data Archive \footnote{https://heasarc.gsfc.nasa.gov} from August 2008 to July 2018, including 11 sGRBs and 104 lGRBs. Notably, previous works mainly focused on the Swift/BAT GRB samples (e.g. \citealt{2015ApJS..218...13Y,2023ApJ...958...37D}, P16, D22). Some relevant parameters are presented in Table~\ref{tab:long}, in which includes GRB name (Column 1), redshift $z$ (Column 2), $T_{90}$  (Column 3), peak energy flux  (Column 4) and K-correction factor (Column 5) for luminosity $L$ (Column 6).


It is worth noting that the K-correction has been made to convert the observed luminosities into the bolometric luminosities within $1-10^{4}$ keV band in the rest frame \cite[e.g.][]{2018PASP..130e4202Z}. Here, the bolometric luminosity reads $L=4\pi d_{\rm L} ^{2}(z) FK$, where $d_{\rm L}$ is the luminosity distance determined by $d_{\rm L}(z)=\frac{c}{H_{\rm 0} } (1+z) \int_{0}^{z} \frac{dz}{\sqrt{(1+z)^{3} \Omega _{\rm M} +\Omega _{\rm \Lambda } } }$ for a flat $\Lambda \rm CDM$ Universe. The cosmological parameters of $H_{\rm 0}  =70 \ \rm km \ \rm s^{-1} \rm Mpc^{-1} $, $\Omega _{\rm M}=0.27 $ and $\Omega _{\rm \Lambda } =0.73$ have been assumed throughout the paper.

\section{Method}\label{method}

\begin{figure*}
     \centering
     \includegraphics[width=0.75\textwidth]{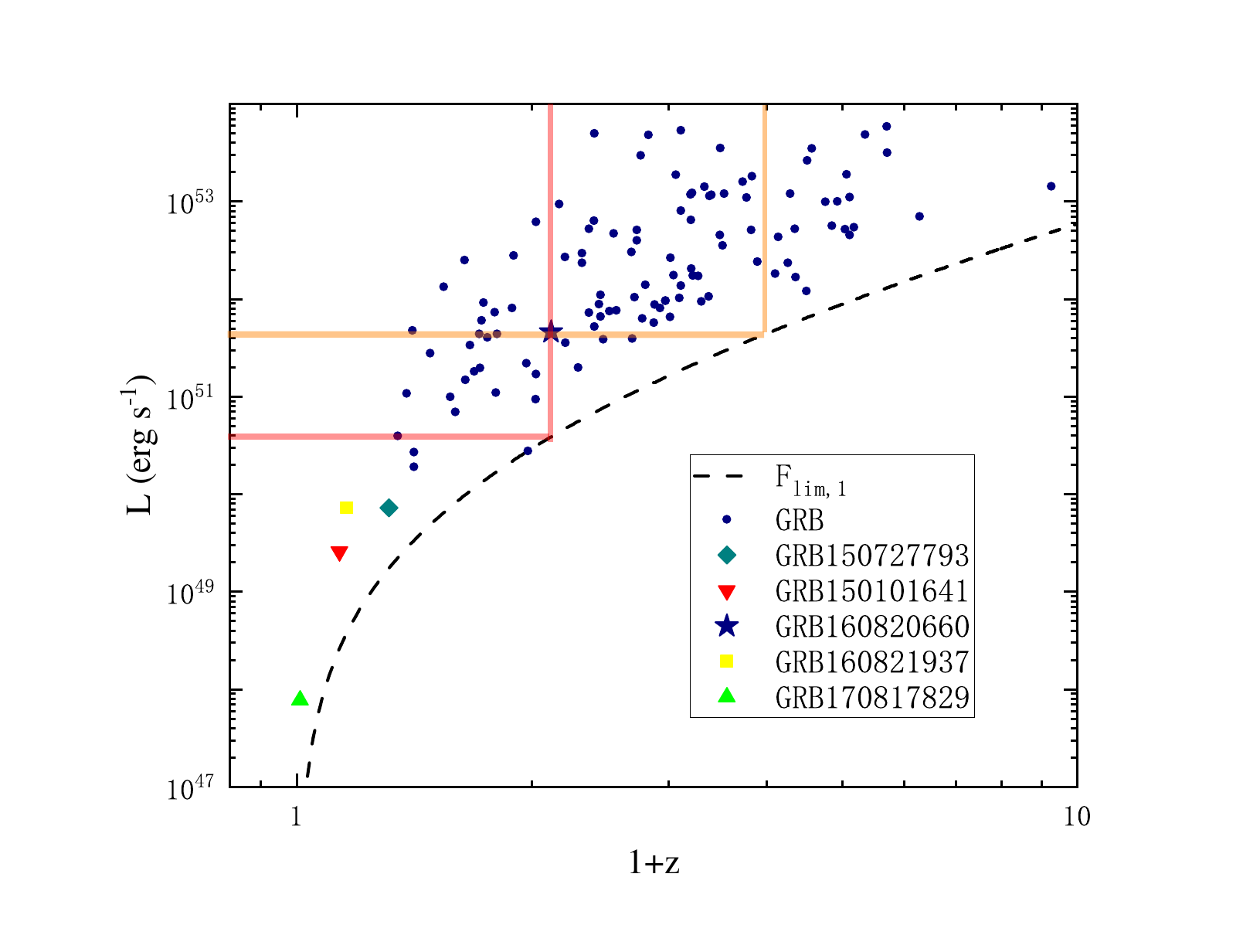}
	\caption{Luminosity is plotted against redshift for 115 Fermi/GBM GRBs. The dashed line represents the lower limit of luminosities constrained by the instrumental sensitivity of $F_{\rm lim,2}$ = $\mathrm{2\times 10^{-8} erg \ cm^{-2}\ s^{-1}}$. Several peculiar GRBs are marked in the insert.  \label{fig:1}}
\end{figure*}

As presented in Figure \ref{fig:1}, the luminosity is positively correlated with the redshift for our sample of 115 Fermi/GBM GRBs. We adopt a power-law form of $g(z) = (1+z)^{k} $, simliar to many previous researchers \citep{2015ApJS..218...13Y, 2015ApJ...800...31D, 2018ApJ...852....1Z, 2022MNRAS.513.1078D}, to depict the effect of redshift evolution. Thus, the luminosity function can be degenerated into $\Psi (L,z) = \psi (L/g(z))\phi (z)$. Letting $L_{0} = L/g(z)$, one can rewrite the function to be $\Psi (L_{0} ,z) = \psi (L_{0} )\phi (z)$, in which the $L_0$ is independent of $z$. For the non-parametric $\tau$ statistic test, assuming a $(z_i, L_{0,i})$ data set, we define $J_i$ data set, of which the number is $n_i$, as 
\begin{equation}\label{eq5}
	J_{i} = \left \{ j \mid L_{0,j} \ge L_{0,i},z_{j}\le z_{i}^{max}   \right \},
\end{equation} where $L_0,i$ is the redshift-de-evolved luminosity of GRBs and $z_{i}^{max}$ represents the maximum redshift that can be detected by the Fermi/GBM. After the $i{\rm th}$ GRB removed from the $J_i$, a new data set, named $J_{i}^{'}$ , can be written as

\begin{equation}\label{eq6}
	J_{i}^{'}  = \left \{ j \mid L_{0,j} \ge L_{0,i}^{min} ,z_{j}\le z_{i}   \right \},
\end{equation} where $L_{i,0} ^ {min}$ is the minimum luminosity of a GRB with $z_i$ that can be detected by Fermi/GBM. And the number in set $J_{i}^{'}$ defined as $M_{i}$. The $\tau$ value can be written as
\begin{equation}\label{eq7} 
\tau \equiv \frac{\sum _{i} \left ( R_{i} - E_{i}  \right ) }{\sqrt{\sum _{i} V_{i} } } ,
\end{equation} 
in which $R_{i}$ is the number of GRBs with redshift $z$ less than or equal to $z_{i}$, and $E_{i} = (1+n_{i})/2 $ and $V_{i} =(n_{i}-1)^2/12$ are the expected mean value and the variance of $R_{i}$. In principle, here $R_{i}$ should be uniformly distributed between 1 and $n_{i}$, which means the sample number of $R_{i} \ge E_{i}$ should be as close as possible to that of $R_{i} <  E_{i}$ and the $\tau$ value should be nearly zero, and then both $L_{0}$ and $z$ would be two irrelevant variables. To achieve this condition, we need to adjust the values of $k$, which should vary with different samples and thresholds of detectors.

Now, we apply the Lynden-Bell's c$^{-}$ method to determine the luminosity function and event rate of GRBs. The cumulative luminosity and redshift distributions can be respectively expressed as
\begin{equation}\label{eq8}
	\Psi (L_{0i} ) = \prod_{j<i}^{} (1+\frac{1}{N_{j} } ),
\end{equation}
and
 \begin{equation}\label{eq9}
 	\phi (z_{i} ) = \prod_{j<i}^{} (1+\frac{1}{M_{j} } ),
 \end{equation}
where $N_i=n_i-1$ and $M_{i}$ are the GRB numbers within the $J_i$ and $J^{'} _{i}$ sets, respectively. The GRB event rate can be estimated by

\begin{equation}\label{eq10}
	\rho (z) = \frac{d\phi (z)}{dz}(1+z) (\frac{dV(z)}{dz})^{-1} ,
\end{equation}
where the differential comoving volume is determined by
\begin{equation}\label{eq11}
	\frac{dV(z)}{dz} = \frac{c}{H_{0} } \frac{4\pi d_{L}^{2} (z)}{(1+z)^{2} } \frac{1}{\sqrt{\Omega _{\Lambda } +\Omega _{M}(1+z)^{3} } } .
\end{equation}

\section{Results} \label{sec:three} 
\subsection{The instrument effect}
\begin{figure*}
	\centering
	\includegraphics[width=0.75\textwidth]{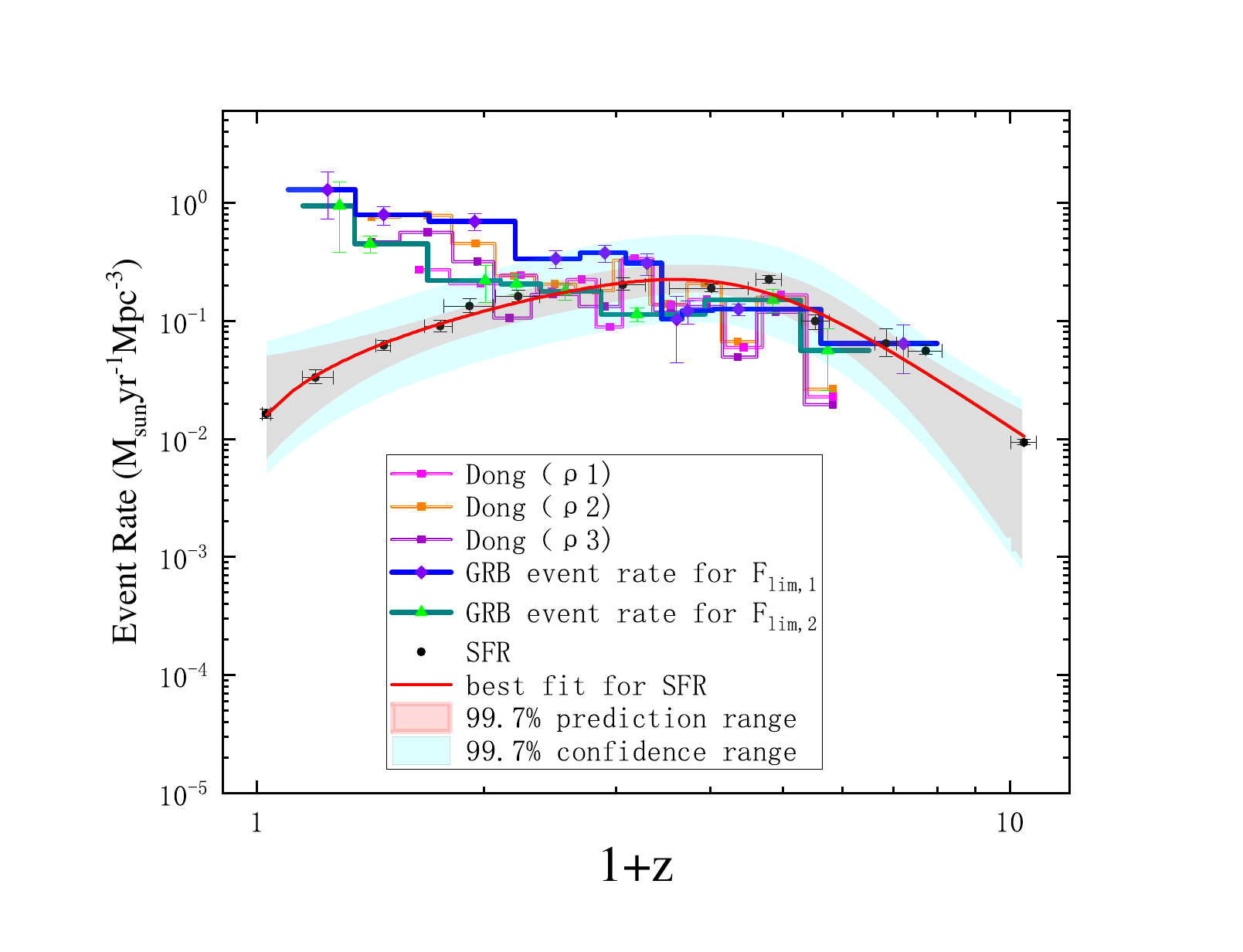}
	\caption{ Comparison of the GRB event rates with the SFR. The blue and green step lines are the GRB rates given by the sensitivities of $F_{\rm lim,1}$ and $F_{\rm lim,2}$, respectively. The magenta, purple and orange step lines denote the Swift lGRB rates for diverse samples in D22. The black circles and the solid curve stand for the average SFR in each bin and the best fitting curve with 1$\sigma$ and 3$\sigma$ confidence levels. \label{fig:2}}
\end{figure*}

To check how the sensitivity affects the deduced event rate of GRBs, we choose two flux cuts of ${F_{\rm lim,1} = 1\times 10^{-7} \rm erg \ \rm cm^{-2}\ \rm s^{-1}}$ and  ${F_{\rm lim,2} = 2\times 10^{-8} \rm erg\ \rm cm^{-2}\ \rm s^{-1}}$ in \cite{2022MNRAS.513.1078D} to calculate the luminosity limits for the sample selection. Then, we apply the non-parametric method as illustrated in Section \ref{method} to infer the event rates of two flux-limited GRB samples and compare with the SFR, of which the SFR data have been collected from the literature \citep{2004ApJ...615..209H, 2006ApJ...647..787T, 2008MNRAS.388.1487L,  2011Natur.469..504B}. After separating the SFR observations into distinct bins, we obtain 17 binned SFR data points as displayed in Figure \ref{fig:2}. Then we fit the SFR data with the empirical model of $\rho _{*}  = (a+bz)h/\left [ 1+(z/c)^{d} \right ] $ \citep{2006ApJ...651..142H} and obtain the best fitted parameters of $a=0.12\pm 0.02, b=0.14\pm 0.02, c=4.61\pm 0.20, d=5.55\pm 0.23$ and $h=0.7$. In addition, we also compare our results with Swift lGRB rates given by D22 for different samples in Figure \ref{fig:2}, where we find that GRB rates detected by diverse satellites exceed the SFR at lower redshift ($z<1$) in a similar way. This means that the observed GRB rates are indepedent of both the energy bands and the flux thresholds of detectors.

\subsection{The selection effect} 
\begin{figure*}
     \centering
     \includegraphics[width=0.75\textwidth]{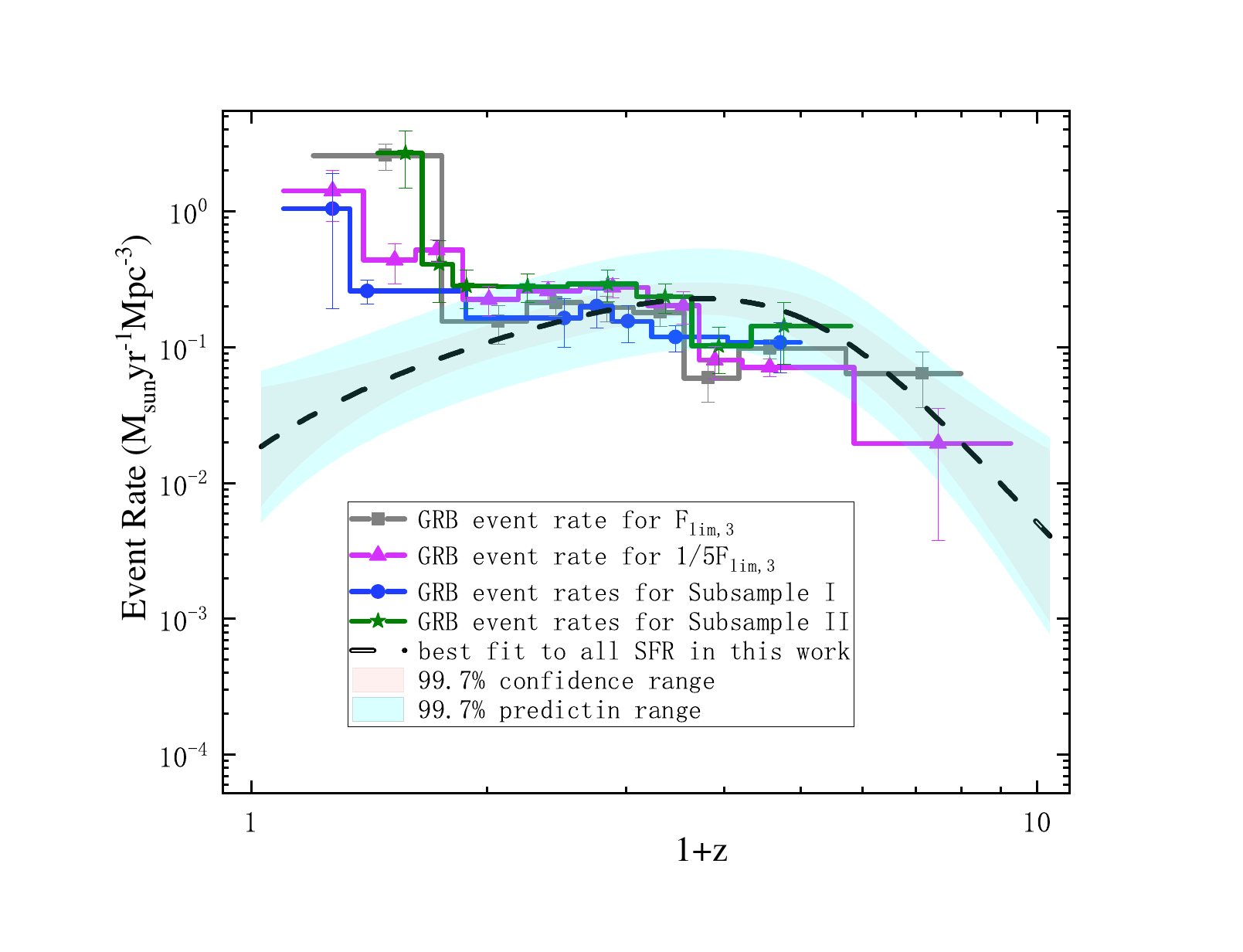}
	\caption{Comparison of the GRB rate versus the SFR between P16 and this work. The gray step line is the observed GRB rate given by the $F_{\rm lim,3}$ of P16, the magenta step line  shows the GRB rate in the case of $F_{\rm lim,3}/5$. The blue and  green step lines are the GRB rates for Subsample I and II, correspondingly. The black dashed line represents the best fitting curve with 1$\sigma$ and 3$\sigma$ confidence levels.  \label{fig:3}}
\end{figure*}

\begin{figure*}
     \centering
     \includegraphics[width=0.75\textwidth]{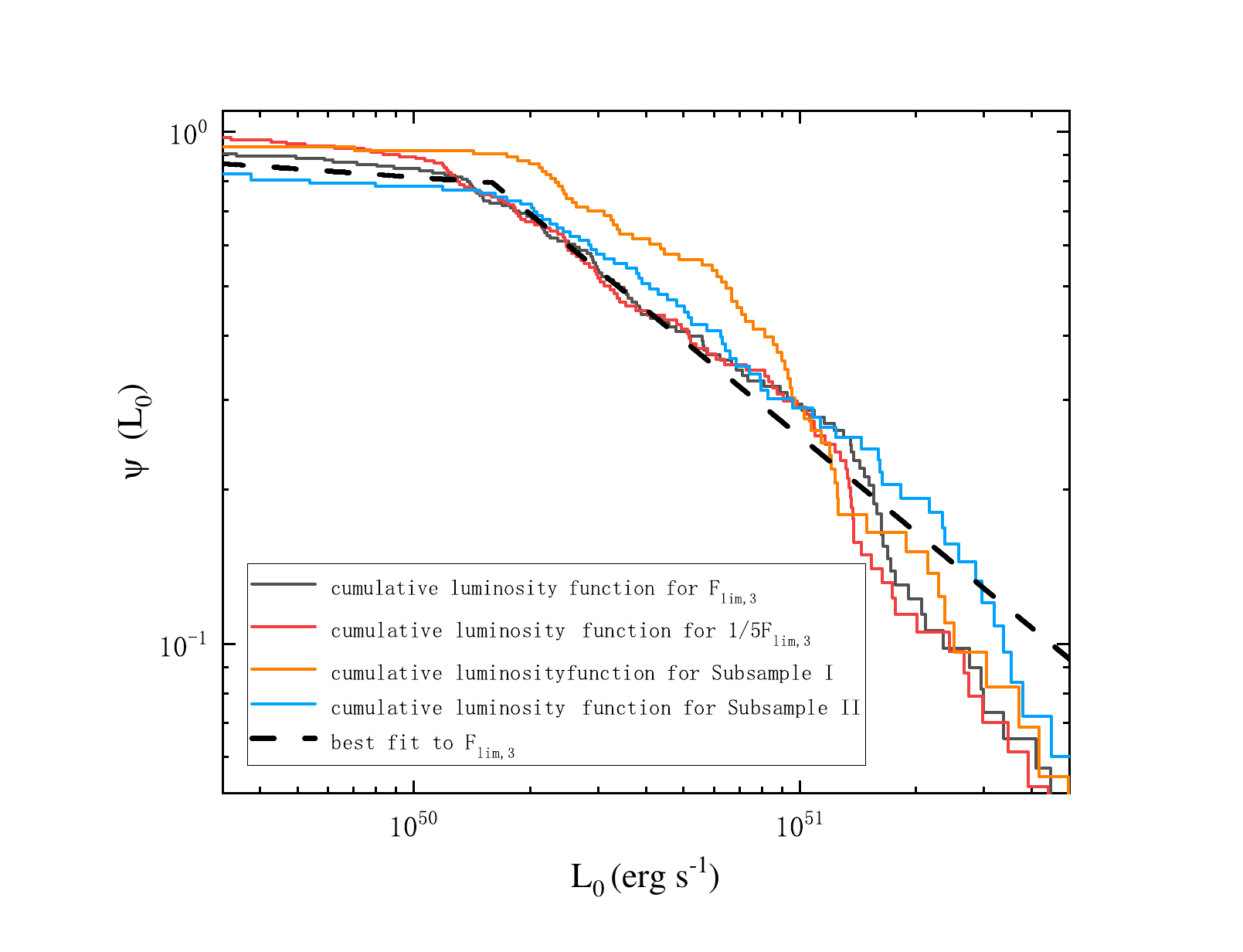}
	\caption{Luminosity distributions of diverse GRB samples. The dashed line shows the best fit with a broken power-law form to the luminosity function constrained by the $F_{\rm  lim,3}$ in P16.
    \label{fig:4}}
\end{figure*}

It can be realized in Section 4.1 that the GRB rates do not depend on the sample selection too much even though different flux cuts have been taken for the truncated data. Besides, sample completeness degree as another selection effect would bias the inferred GRB rate more or less and will be paied particular attention. Previously, P16 pointed out that the incompleteness of GRB samples would cause the low-redshift excess of GRB rate compared with the SFR. In fact, the low-redshift excess had been announced for the incomplete GRB samples in many works. Recently, D22 verified with diverse samples that the low-redshift excess is intrinsic and does not rely on whether the GRB sample is complete or not. Particularly, they found that the excess of the complete sample built with the selection conditions \citep{2006A&A...460L..13J, 2012ApJ...749...68S} at lower redshift becomes more prominent, which is opposite to the conclusion in P16. Now, we choose the same flux limit of $\mathrm{F_{lim,3} = 5\times 10^{-8} erg \ cm^{-2}\ s^{-1}}$ as P16 and calculate the event rate of Fermi/GBM GRBs. Simultanously, a flux threshold of 1/5 $F_{\rm lim,3}$ is also tested. Figure \ref{fig:3} shows that the event rates of two above samples evolve with redshift similarly and exceed the SFR at low redshift identically. Furthermore, we adopt a random sampling program to remove those GRBs around the the luminosity limit to consitute our sample I of 71 GRBs. We also implement the random program to pick the GRBs above the luminosity limit to comprise our subsample II with the same number. The reason why both subsamples are set to be equal is aimed to reduce the bias of sample size. Using the Lynden-Bell's c$^-$ method, we estimate the GRB event rates of subsamples I and II and plot both evolution histories in Figure \ref{fig:3}, from which we can see that the GRB event rates of four differnt samples always exceed the SFR at low-redshift of $z<1$, regardless of the sample completeness. Cumulative de-evolved luminosity distributions are compared in Figure \ref{fig:4}, where four samples show the comparable characteristic of luminosity distribution. A broken power-law function has been employed to fit the luminosity distribution established on $F_{\rm lim,3}$ and returns two indices of $a_{1}=-0.05\pm0.01$ and $a_{2}=-0.62\pm0.01$ together with a broken luminosity at $L_{\rm b} =1.6\times10^{50}$ erg $s^{-1}$. Note that the $k$ values corresponding to $F_{\rm lim,1}$, $F_{\rm lim,2}$, $F_{\rm lim,3}$, $F_{\rm lim,3}/5$, subsamples I and II are respectively 3.29, 3.85, 3.83, 3.93, 3.69 and 3.83 according to the $\tau$ statistics \citep{1992ApJ...399..345E}.

\subsection{Low-redshift excess of GRB rate}
\begin{figure*}
     \centering
     \includegraphics[width=0.75\textwidth]{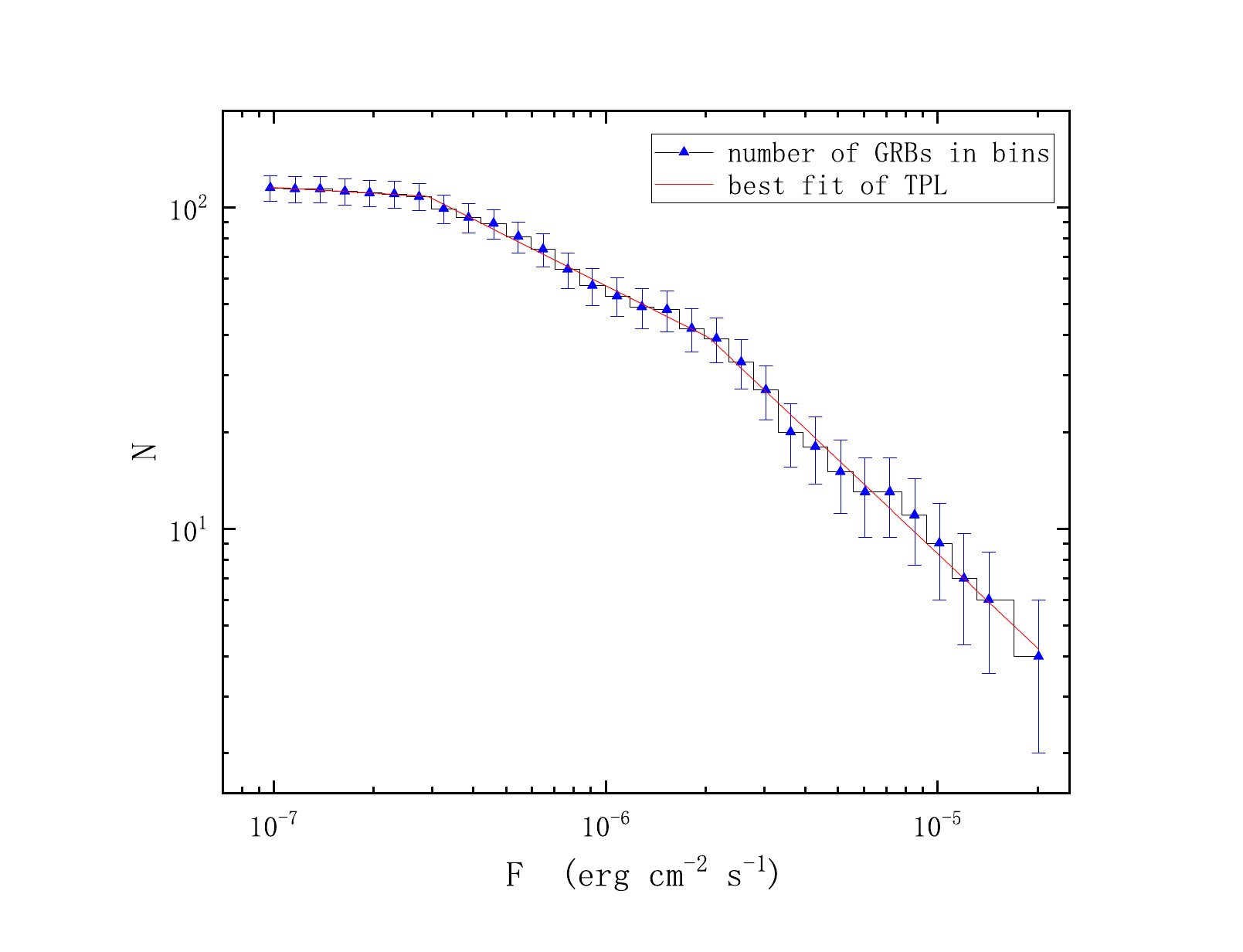}
	\caption{The cumulative peak flux distribution of 115 Fermi/GBM GRBs. The solid line is the best fit with a TPL function ($\chi_{\nu}^2\approx1.51$). \label{fig:5}}
\end{figure*}

\begin{figure*}
     \centering
     \includegraphics[width=0.75\textwidth]{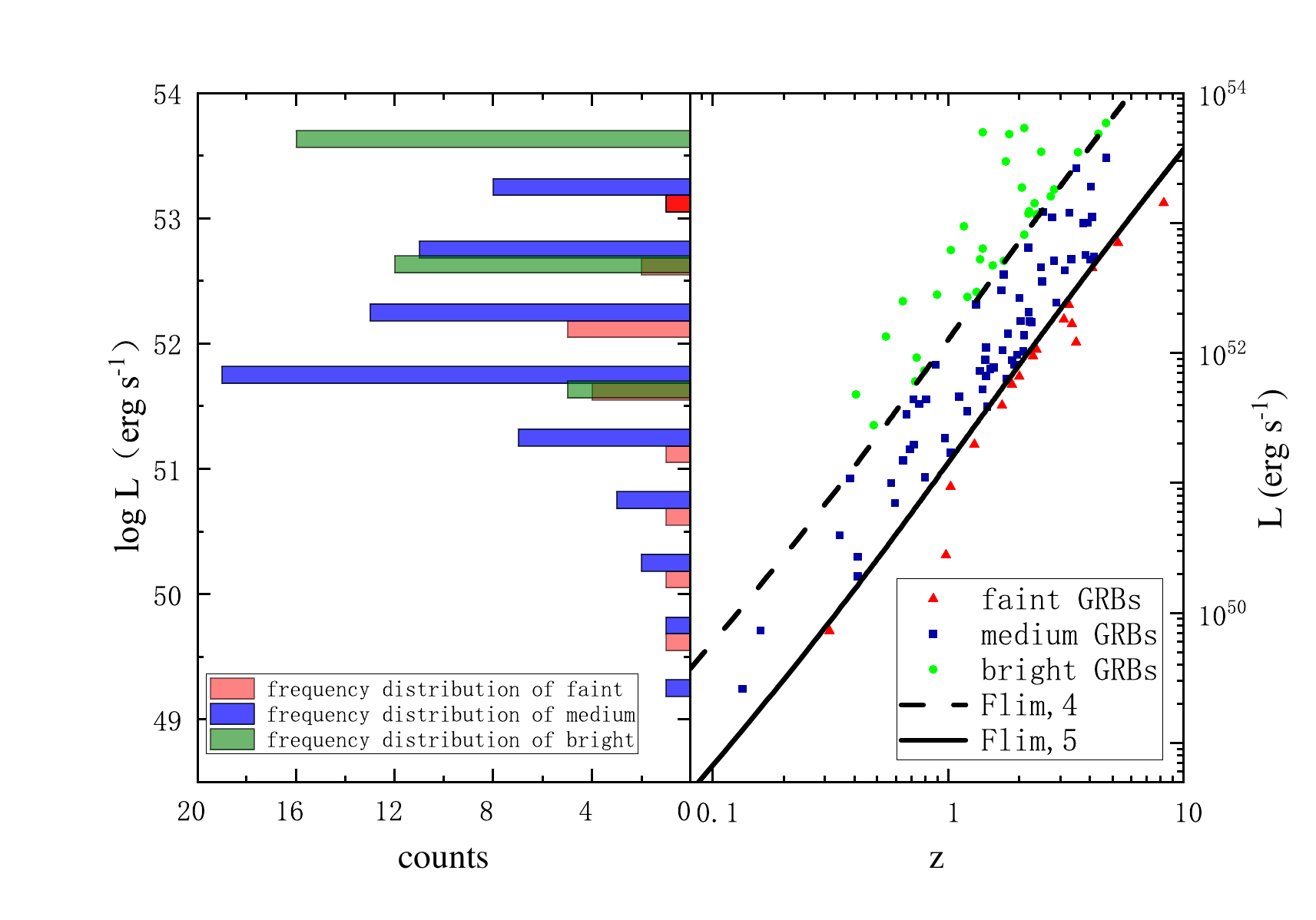}
	\caption{ Left panel:  luminosity histograms of the bright, medium and faint GRBs. Right panel:  luminosity versus redshift for three kinds of GRBs. Two luminosity thresholds estimated by $F_{\rm {lim,4}}$ and $F_{\rm {lim,5}}$ are drawn by the dashed and solid lines in each. \label{fig:6}}
\end{figure*}

To gain the reliable estimate of GRB rate, some authors usually insisted that the selected GRBs should comply with the completeness conditions \citep{2006A&A...460L..13J, 2012ApJ...749...68S} and consist of bright GRBs with peak flux higher than 2.6 ph $\rm cm^{2}$ $\rm s^{-1}$ \cite[see e.g.][]{2016A&A...587A..40P}. At present, we draw the flux distribution in Figure \ref{fig:5} and fit it with a triple power-law (TPL) function \citep[see Eq. (17) in][]{2015ApJ...812...33S} 
and divide our GRB sample into three classes, namely bright, medium and faint ones.
Figure \ref{fig:5} illustrates the best fit of the cumulative flux distribution with two broken fluxes of $\mathrm{F_{b,1} =(2.05\pm1.69)\times 10^{-6} \ erg \ cm^{-2} \ s^{-1}}$ and $\mathrm{F_{b,2} =(2.91\pm0.17)\times 10^{-7} \ erg \ cm^{-2} \ s^{-1}}$ ($\chi_{\nu}^2\approx$1.51), which are utilized to regroup our sample into 34 bright, 65 medium, and 16 faint GRBs. For the bright and medium bursts, we set the flux cuts to be $F_{\rm lim,4}=2.35\times 10^{-6} \rm erg\ \rm cm^{-2} \rm s^{-1}$ and
$F_{\rm lim,5}=2.65\times 10^{-7} \rm erg\ \rm cm^{-2} \rm s^{-1}$, respectively, which have been employed to calculate the luminosity limits as displayed in right panel of Figure \ref{fig:6}. The luminosity histograms of three kinds of GRBs are shown in left panel of Figure \ref{fig:6}, in which we notice that the bright GRB sample has the highest luminosity on average.

\begin{figure*}
     \centering
     \includegraphics[width=0.75\textwidth]{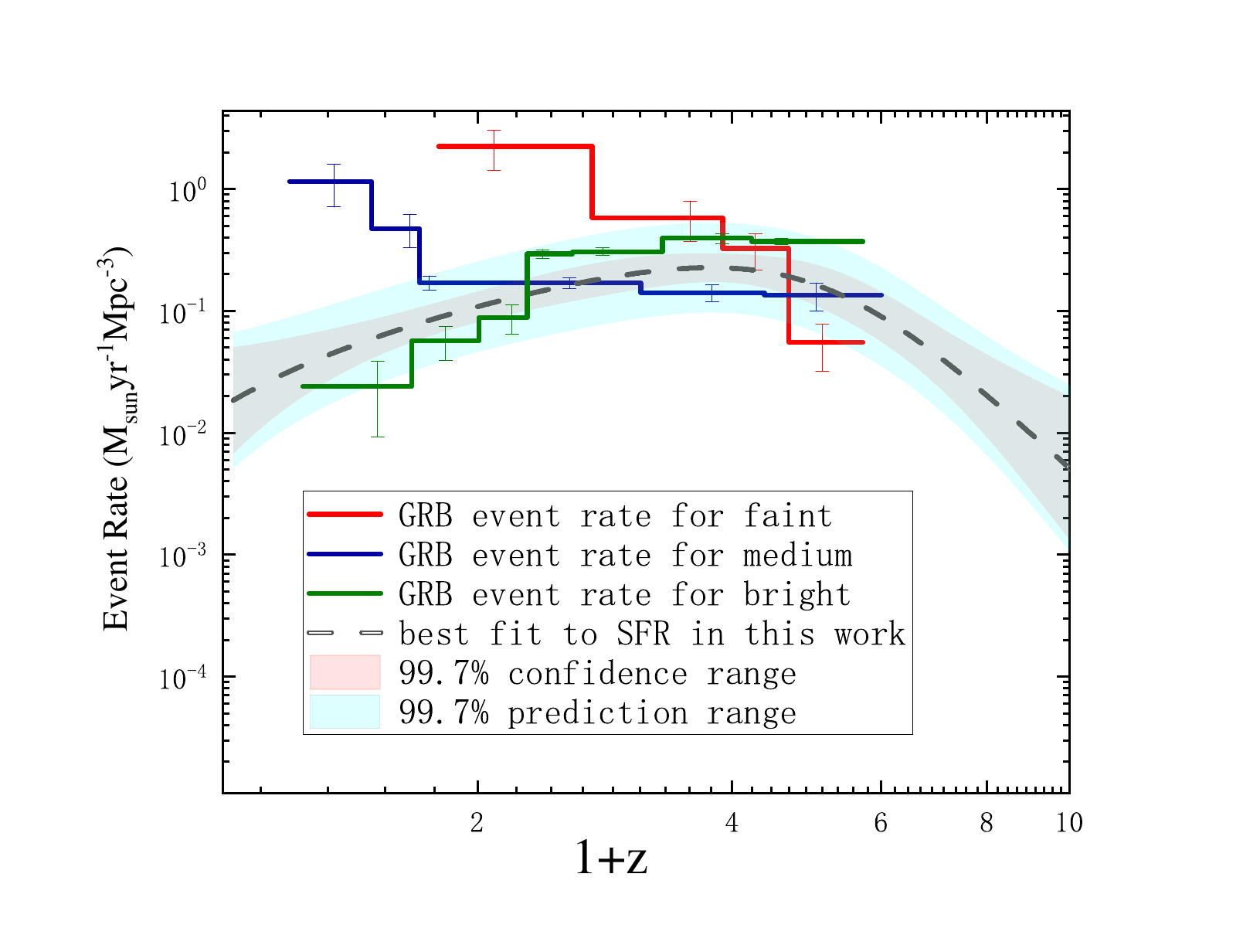}
	\caption{Comparison of the event rate with the SFR for three GRBs categories of this work. The estimated event rates of bright, medium and faint GRBs are marked by the green, blue and red step lines, respectively. The black dashed line is the best fitted curve together with  confidence levels of $1\sigma$ and $3\sigma$ shown respectively by the gray and light blue regions.\label{fig:7}}
\end{figure*}

By adopting the above non-parametric method, we compute the event rates of three kinds of GRBs and compare them with the SFR in Figure \ref{fig:7}, from which we find that the event rates of faint and medium GRBs exceed the SFR accordantly, while the bright GRB rate matches the SFR well. We argue that the excess of the GRB rates at low-redshift is contributed by the faint and medium GRBs rather than the bright ones. It can be understood that these bright GRBs hold higher luminosities and are mainly produced from the core-collapse of massive stars as proposed by D22. This implies that most of the faint and bright GRBs might have different physical origins in theory \citep{2024ApJ...963L..12P,2024MNRAS.535.2800L}. Interestingly, we run a two-dimensional K-S test to any two distributions and find that the faint and medium GRBs, unlike  bright bursts, are identically distributed. Note that the $k$ values in the $\tau$ statistic test are estimated to be about 0.13, 3.89 and 3.5 for bright, medium and faint GRBs, respectively, which manifests the luminosities of bright GRBs weakly depend on the redshift, hinting that a volume-limited GRB sample \citep[see][for FRBs]{2025A&A...698A..18Z} could be better for reckoning the GRB event rates.

\section{Summary} \label{sec:four}

In this work, we have comparatively studied the effects of sample selection, instrumental threshold, energy bands, sample completeness and source brightness of Fermi/GBM bursts on the relationships between the GRB rate and the SFR. The $\tau$ statistics and the Lynden-Bell’s c$^-$ method are applied to estimate the GRB event rates. We can draw the following conclusions:
\begin{itemize}
\item We verified that Fermi/GBM GRBs exhibit higher event rate than the SFR at lower redshift of $z<1$ and match with each other at higher redshift despite different flux sensitivities considered, which is in good agreement with the findings for Swift lGRBs in D22. It demonstrates that the low-redshift excess of GRB rate is not related with the differences in either energy bands or flux limits.
\item Regardless of the sample completeness, we adopt the same flux thresholds as P16 to build different subsamples of Fermi GRBs and calculate their corresponding event rates. It is interestingly found that all these formation rate histories compared with the SFR show an obvious excess at lower redshift of $z<1$ and they evolve with redshift uniformly, which is in agreement with previous studies. 

\item In order to interpret the low-redshift excess of GRB rate in contrast to the SFR, we have divided 115 Fermi GRBs into three groups, namely bright, medium and faint GRBs. We surprisingly notice that the deduced GRB rates of medium and faint GRBs exceed the SFR at lower redshift, while the bright GRB rate follows the SFR perfectly. This is attributed to the fact that these bright GRBs possess relatively high luminosity and show weak dependence of luminosity on redshift. 

\end{itemize}
\section{Acknowledgements} \label{sec:seven}

We would like to thank the anonymous referee for very valuable comments and suggestions that lead to an overall improvement of this study. This work was supported in part by National Natural Science Foundation of China (grant No. U2031118).
\section*{Data Availability Statement}
The Fermi/GBM data used here can be retrieved from the official Fermi satellite (https://fermi.gsfc.nasa.gov/ssc/data/). Partial data of Swift GRB rates can be referred to  \cite{2022MNRAS.513.1078D,2023ApJ...958...37D}.


\clearpage
\onecolumn
\begin{longtable}{p{3cm}<{\centering}p{1.5cm}<{\centering}p{1.5cm}<{\centering}p{3cm}<{\centering}p{1.5cm}<{\centering}p{3cm}<{\centering}}
    \caption{Typical parameters of 115 selected Fermi/GBM GRBs.} 
    \label{tab:long} \\
    \hline
    \hline
    \multicolumn{1}{c}{Name} & 
    \multicolumn{1}{c}{$z$} & 
    \multicolumn{1}{c}{$T_{90}$} & 
    \multicolumn{1}{c}{$\text{Peak Flux}$} & 
    \multicolumn{1}{c}{$K$} & 
    \multicolumn{1}{c}{$\text{Bolometric luminosity}$} \\
    \multicolumn{1}{c}{\textbf{}} & 
    \multicolumn{1}{c}{\textbf{}} & 
    \multicolumn{1}{c}{\textbf{($\mathrm{s}$)}} & 
    \multicolumn{1}{c}{\textbf{($\mathrm{erg \, cm^{-2}\, s^{-1}}$)}} & 
    \multicolumn{1}{c}{\textbf{}} & 
    \multicolumn{1}{c}{\textbf{($\mathrm{erg\, s^{-1}}$)}} \\
    \hline
    \endfirsthead
    \multicolumn{6}{c}{{\bfseries \tablename\ \thetable{} -- continued from previous page}} \\
    \hline
    \multicolumn{1}{c}{Name} & 
    \multicolumn{1}{c}{$z$} & 
    \multicolumn{1}{c}{$T_{90}$} & 
    \multicolumn{1}{c}{$\text{Peak Flux}$} & 
    \multicolumn{1}{c}{$K$} & 
    \multicolumn{1}{c}{$\text{Bolometric luminosity}$} \\
    \multicolumn{1}{c}{\textbf{}} & 
    \multicolumn{1}{c}{\textbf{}} & 
    \multicolumn{1}{c}{\textbf{($\mathrm{s}$)}} & 
    \multicolumn{1}{c}{\textbf{($\mathrm{erg \, cm^{-2}\, s^{-1}}$)}} & 
    \multicolumn{1}{c}{\textbf{}} & 
    \multicolumn{1}{c}{\textbf{($\mathrm{erg\, s^{-1}}$)}} \\
    \hline
    \endhead
    \hline
    \endfoot
    \hline
    \endlastfoot
GRB080804972 &  2.2045 &  24.704 & (36.1 $\pm$ 2.8)$\times 10^{-7}$ & 0.95  & (77.0 $\pm$ 6.1)$\times 10^{46}$   \\
GRB080810549$^{\ast }$ &  3.35  & 75.201& (54.0$\pm$ 7.8)$\times 10^{-8}$  & 1.42  & (25.9$\pm$ 3.7)$\times 10^{48}$  \\
GRB080905705$^{\$ }$  &  2.374 & 105.984 & (10.3$\pm$ 2.8)$\times 10^{-7}$  & 1.01 & (7.3$\pm$ 2.0)$\times 10^{49}$  \\
GRB080916009 &    4.35  & 62.977 & (22.4$\pm$ 3.2)$\times 10^{-8}$  & 0.60 & (7.2$\pm$ 1.1)$\times 10^{49}$  \\
GRB080916406 & 0.689 & 46.337   & (95.9$\pm$ 7.1)$\times 10^{-8}$  & 1.22  & (39.4$\pm$ 2.9)$\times 10^{49}$  \\
GRB080928628 & 1.692  &   14.336  & (20.7$\pm$ 1.4)$\times 10^{-7}$  & 0.86 & (108.0$\pm$ 7.2)$\times 10^{49}$ \\
GRB081008832 & 1.9685 &126.722  & (80.0$\pm$ 5.8)$\times 10^{-7}$  & 0.85 & (47.9$\pm$ 3.5)$\times 10^{50}$  \\
GRB081109293 &  0.9787 &  58.369 & (30.6$\pm$ 6.2)$\times 10^{-8}$  & 0.42 & (19.1$\pm$ 3.8)$\times 10^{49}$  \\
GRB081121858 & 2.512  &   41.985 & (43.3$\pm$ 6.5)$\times 10^{-8}$  & 0.67 & (27.0$\pm$ 4.1)$\times 10^{49}$  \\
GRB081222204 & 2.77    & 18.88   & (30.7$\pm$ 6.6)$\times 10^{-7}$  & 1.00  & (27.8$\pm$ 6.0)$\times 10^{50}$  \\
GRB090102122 & 1.547   & 26.624  & (111.0$\pm$ 1.4)$\times 10^{-7}$ & 0.99 & (134.0$\pm$ 1.7)$\times 10^{50}$ \\
GRB090323002 & 3.57    & 133.89  & (72.6$\pm$ 3.9)$\times 10^{-8}$  & 1.08 & (99.2$\pm$ 5.4)$\times 10^{49}$  \\
GRB090328401 & 0.736   & 61.697  & (46.1$\pm$ 5.1)$\times 10^{-8}$  & 0.77 & (69.7$\pm$ 7.6)$\times 10^{49}$  \\
GRB090423330 & 8.26    & 7.168   & (138.0$\pm$ 1.5)$\times 10^{-7}$ & 0.94 & (250.0$\pm$ 2.6)$\times 10^{50}$ \\
GRB090424592 $^{\$ }$ & 0.544   & 14.144  & (8.2$\pm$ 1.2)$\times 10^{-7}$  & 1.03  & (14.9$\pm$ 2.2)$\times 10^{50}$  \\
GRB090516137 & 4.109   & 110.594 & (169.0$\pm$ 6.0)$\times 10^{-8}$ & 0.64& (33.7$\pm$ 1.19)$\times 10^{50}$  \\
GRB090516353 & 4.109   & 123.138 & (84.4$\pm$ 8.9)$\times 10^{-8}$  & 0.96 & (18.1$\pm$ 1.9)$\times 10^{50}$  \\
GRB090519881 & 3.85 & 74.177  & (186.0$\pm$ 6.7)$\times 10^{-8}$ & 1.06  & (43.8$\pm$ 1.6)$\times 10^{50}$  \\
GRB090902462 & 1.822   & 19.328  & (8.3$\pm$ 2.2)$\times 10^{-7}$  & 0.87 & (19.7$\pm$ 5.1)$\times 10^{50}$  \\
GRB090926181 & 2.1062  & 13.76   & (24.8$\pm$ 1.3)$\times 10^{-7}$  & 1.01  & (60.5$\pm$ 3.2)$\times 10^{50}$  \\
GRB090926914 & 2.1062  & 64.001  & (36.5$\pm$ 1.0)$\times 10^{-7}$  & 0.94& (92.3$\pm$ 2.6)$\times 10^{50}$  \\
GRB090927422 & 1.37    & 0.512   & (150.0$\pm$ 6.5)$\times 10^{-8}$ & 0.76 & (40.3$\pm$ 1.8)$\times 10^{50}$  \\
GRB091003191 & 0.8969  & 20.224  & (24.0$\pm$ 4.0)$\times 10^{-7}$  & 0.98 & (7.3$\pm$ 1.2)$\times 10^{51}$  \\
GRB091020900 & 1.71    & 24.256  & (35.2$\pm$ 3.4)$\times 10^{-8}$  & 0.50& (11.0$\pm$ 1.1)$\times 10^{50}$  \\
GRB100206563 & 0.4068  & 0.176   & (138.0$\pm$ 5.9)$\times 10^{-8}$ & 1.09 & (43.8$\pm$ 1.9) $\times 10^{50}$  \\
GRB100414097 & 1.368   & 26.497  & (20.0$\pm$ 1.2)$\times 10^{-7}$  & 0.98 & (80.8$\pm$ 4.8)$\times 10^{50}$  \\
GRB100728095 & 2.106   & 165.378 & (67.9$\pm$ 1.3)$\times 10^{-7}$  & 0.82& (281.0$\pm$ 5.2)$\times 10^{50}$  \\
GRB100728439$ ^{\$}$ & 1.567   & 10.24   & (4.37 $\pm$ 1.2)$\times  10^{-7} $ & 1.10& (22.0 $\pm$ 5.9)$\times 10^{50}$ \\
GRB100814160 & 1.44    & 150.53  & (54.1$\pm$ 9.2)$\times 10^{-9}$  & 0.17& $(27.8\pm 4.7)\times 10^{49}$ \\
GRB100906576$^{\$}$  & 1.727   & 110.594 & (16.4$\pm$ 3.5)$\times 10^{-8}$  & 1.10& (9.4$\pm$ 2.0)$\times 10^{50}$ \\
GRB101213451 & 0.414   & 5.12    & (29.3$\pm$ 8.2)$\times 10^{-8}$  & 1.00 & (17.0$\pm$ 4.7)$\times 10^{50}$ \\
GRB101213849 & 0.414   & 45.057  & (106.0$\pm$ 1.9)$\times 10^{-7}$ & 0.87 & (61.8$\pm$ 1.10)$\times 10^{51}$  \\
GRB110205027 & 2.22    & 5.376   & (64.3$\pm$ 5.1)$\times 10^{-8}$  & 0.62 & (46.0$\pm$ 3.7)$\times 10^{50}$ \\
GRB110731465 & 2.83    & 7.485   & (117.0$\pm$ 1.2)$\times 10^{-7}$ & 0.79 & (942.0$\pm$ 9.9)$\times 10^{50}$\\
GRB110818860 & 3.36    & 67.073  & (310.0$\pm$ 7.9)$\times 10^{-8}$ & 1.05 & (270.0$\pm$ 6.9)$\times 10^{50}$\\
GRB111107035 & 2.893   & 12.032  & (40.6$\pm$ 5.5)$\times 10^{-8}$  & 1.01 & (35.5$\pm$ 4.8)$\times 10^{50}$ \\
GRB111117510 $^{\ast }$ & 2.211   & 0.432   & (19.1$\pm$ 3.7)$\times 10^{-8}$  & 0.60 & (19.8$\pm$ 3.8)$\times 10^{50}$ \\
GRB111228657$^{\star\$}$  & 0.714   & 99.842  & (216.0$\pm$ 9.4)$\times 10^{-8}$ & 1.00 & (23.5$\pm$ 1.00)$\times 10^{51}$  \\
GRB120119170 & 1.728   & 55.297  & (271.0$\pm$ 8.3)$\times 10^{-8}$ & 1.03 & (295.0$\pm$ 9.0)$\times 10^{50}$\\
GRB120326056 & 1.798   & 11.776  & (44.3$\pm$ 1.1)$\times 10^{-7}$  & 0.84& (52.5$\pm$ 1.3)$\times 10^{51}$  \\
GRB120624933 & 2.1974  & 271.364 & (6.11$\pm$ 1.6)$\times 10^{-7}$  & 1.00  & (7.3$\pm$ 1.9)$\times 10^{51}$  \\
GRB120711115 & 1.405   & 44.033  & (50.2$\pm$ 1.1)$\times 10^{-7}$  & 0.64& (63.7$\pm$ 1.5)$\times 10^{51}$  \\
GRB120712571 & 4.1745  & 22.528  & (41.0$\pm$ 3.9)$\times 10^{-8}$  & 0.63& (52.2$\pm$ 4.9)$\times 10^{50}$ \\
GRB120716712 & 2.486   & 226.048 & (394.0$\pm$ 2.4)$\times 10^{-7}$ & 0.59 & (501.0$\pm$ 3.0)$\times 10^{51}$ \\
GRB120729456$^{\star\$}$ & 0.8     & 25.472  & (65.5$\pm$ 5.7)$\times 10^{-8}$  & 0.61& (88.4$\pm$ 7.7) $ \times 10^{50}$ \\
GRB120907017 & 0.97    & 5.76    & (48.1$\pm$ 7.2)$\times 10^{-8}$  & 1.02 & (66.3$\pm$ 10.0)$\times 10^{50}$ \\
GRB120909070 & 3.93    & 112.066 &(8.0$\pm$ 1.5)$\times 10^{-7}$  & 1.03& (11.0$\pm$ 2.0)$\times 10^{51}$  \\
GRB120922939 & 3.1 & 182.275 & (27.1$\pm$ 4.2)$\times 10^{-8}$  & 0.68 & (38.5$\pm$ 6.0)$\times 10^{50}$ \\
GRB121027038$^{\$}$ & 1.773   & 166.915 & (48.9$\pm$ 5.9)$\times 10^{-8}$  & 1.05  & (75.1$\pm$ 9.1)$\times 10^{50}$ \\
GRB121128212 & 2.2     & 17.344  & (29.1$\pm$ 1.0)$\times 10^{-7}$  & 0.99 & (47.0$\pm$ 1.7)$\times 10^{51}$  \\
GRB121211574 & 1.023   & 5.632   & (46.2$\pm$ 4.9)$\times 10^{-8}$  & 0.66 & (77.0$\pm$ 8.1)$\times 10^{50}$ \\
GRB130215063$^{\star}$ & 0.597   & 143.746 & (151.0$\pm$ 8.3)$\times 10^{-8}$ & 0.97& (30.3$\pm$ 1.7)$\times 10^{51}$  \\
GRB130408653 & 3.757   & 9.216   & (19.6$\pm$ 3.3)$\times 10^{-8}$  & 0.75& (39.5$\pm$ 6.7)$\times 10^{50}$ \\
GRB130505955 & 2.27 & 50.241  & (50.7$\pm$ 3.0)$\times 10^{-8}$  & 0.40 & (105.0$\pm$ 6.17)$\times 10^{50}$\\
GRB130518580 & 2.488   & 48.577  & (18.9$\pm$ 1.4)$\times 10^{-7}$  & 0.82 & (40.1$\pm$ 2.3)$\times 10^{51}$  \\
GRB130610133 & 2.092   & 21.76   & (241.0$\pm$ 7.3)$\times 10^{-8}$ & 0.83& (51.3$\pm$ 1.5)$\times 10^{51}$  \\
GRB130925173 & 0.347   & 215.555 & (134.0$\pm$ 1.6)$\times 10^{-7}$ & 0.84 & (297.0$\pm$ 3.6)$\times 10^{51}$ \\
GRB131004904$^{\$}$ & 0.717   & 1.152   & (27.9$\pm$ 5.1)$\times 10^{-8}$  & 0.60 & (6.3$\pm$ 1.14)$\times 10^{51}$  \\
GRB131011741 & 1.874   & 77.057  & (59.9$\pm$ 5.2)$\times 10^{-8}$  & 1.00& (14.0$\pm$ 1.2)$\times 10^{51}$  \\
GRB131030653 & 1.295   & 53.248  & (199.0$\pm$ 2.0)$\times 10^{-7}$ & 1.00 & (482.0$\pm$ 4.8)$\times 10^{51}$ \\
GRB131105087 & 1.686   & 112.642 & (22.2$\pm$ 8.1)$\times 10^{-8}$  & 0.96& (5.73$\pm$ 2.1)$\times 10^{51}$  \\
GRB131108862 & 2.4 & 18.176  & (33.7$\pm$ 2.4)$\times 10^{-8}$  & 0.40 & (87.6$\pm$ 6.1)$\times 10^{50}$ \\
GRB131117766 & 4.042   & 93.954  & (29.3$\pm$ 6.5)$\times 10^{-8}$  & 1.03 & (8.1$\pm$ 1.8)$\times 10^{51}$  \\
GRB131231198 & 0.642   & 31.232  & (3.29$\pm$ 1.03) $\times 10^{-7}$& 1.07 & (9.65$\pm$ 3.03)$\times 10^{51}$  \\
GRB140206304 & 2.73    & 27.264  & (32.8$\pm$ 7.9)$\times 10^{-8}$  & 0.97 & (9.6$\pm$ 2.3)$\times 10^{51}$  \\
GRB140213807 & 1.2076  & 18.624  & (21.3$\pm$ 3.4)$\times 10^{-8}$  & 0.72 & (6.6$\pm$ 1.0)$\times 10^{51}$  \\
GRB140304557 & 5.283   & 31.232  & (8.5$\pm$ 1.3)$\times 10^{-7}$  & 1.00 & (26.4$\pm$ 4.0)$\times 10^{51}$  \\
GRB140423356 & 3.26    & 95.233  & (54.8$\pm$ 6.0)$\times 10^{-8}$  & 0.90 & (17.6$\pm$ 1.9)$\times 10^{51}$  \\
GRB140428906 & 4.7  & 0.32 & (57.1$\pm$ 1.1)$\times 10^{-7}$  & 0.65 & (188.0$\pm$ 3.5)$\times 10^{51}$ \\
GRB140506880$^{\$}$ & 0.889 & 64.128  & (30.2$\pm$ 7.8)$\times 10^{-8}$  & 1.07 & (10.3$\pm$ 2.7)$\times 10^{51}$  \\
GRB140508128 & 1.027   & 44.288  & (234.0$\pm$ 8.9)$\times 10^{-8}$ & 0.76 & (81.0$\pm$ 3.1)$\times 10^{51}$  \\
GRB140508179 & 1.027   & 19.456  & (39.5$\pm$ 5.7)$\times 10^{-8}$  & 1.04 & (13.7$\pm$ 2.0)$\times 10^{51}$  \\
GRB140512814 & 0.725   & 147.97  & (155.0$\pm$ 1.6)$\times 10^{-7}$ & 0.85 & (539.0$\pm$ 5.5)$\times 10^{51}$ \\
GRB140518709 & 4.707   & 0.704   & (306.0$\pm$ 8.7)$\times 10^{-8}$ & 0.65 & (118.0$\pm$ 3.40$\times 10^{51}$ \\
GRB140606133$^{\star}$ & 0.384   & 22.784  & (167.0$\pm$ 9.9)$\times 10^{-8}$ & 1.00  & (64.6$\pm$ 3.8)$\times 10^{51}$  \\
GRB140620219 & 2.04    & 45.825  & (53.0$\pm$ 8.4)$\times 10^{-8}$  & 1.02 & (20.6$\pm$ 3.3)$\times 10^{51}$  \\
GRB140623224 & 1.92    & 111.104 & (31.4$\pm$ 4.5)$\times 10^{-7}$  & 0.99& (12.3$\pm$ 1.8)$\times 10^{52}$  \\
GRB140703026 & 3.14    & 83.969  & (4.41$\pm$ 1.0)$\times 10^{-7}$  & 0.93 & (17.4$\pm$ 4.0)$\times 10^{51}$  \\
GRB140801792 & 1.32    & 7.168   & (41.3$\pm$ 3.3)$\times 10^{-8}$  & 0.32& (17.2$\pm$ 1.4)$\times 10^{51}$  \\
GRB140808038 & 3.29  & 4.477   & (22.0$\pm$ 4.8)$\times 10^{-8}$  & 0.68 & (9.5$\pm$ 2.1)$\times 10^{51}$  \\
GRB140907672 & 1.21  & 35.841  & (319.0$\pm$ 7.7)$\times 10^{-8}$ & 0.75 & (142.0$\pm$ 3.4)$\times 10^{51}$ \\
GRB141004150$^{\star\$}$ & 0.573   & 9.472   & (22.9$\pm$ 5.9)$\times 10^{-8}$  & 1.00 & (10.6$\pm$ 2.7)$\times 10^{51}$  \\
GRB141028455 & 2.33   & 31.489  & (244.0$\pm$ 7.9)$\times 10^{-8}$ & 0.74 & (114.0$\pm$ 3.7)$\times 10^{51}$ \\
GRB141121414 & 1.47  & 3.84   & (245.0$\pm$ 5.5)$\times 10^{-8}$ & 0.55& (117.0$\pm$ 2.6)$\times 10^{51}$ \\
GRB141220252 & 1.3195  & 7.616   & (87.5$\pm$ 5.3)$\times 10^{-8}$  & 0.80 & (45.6$\pm$ 2.8)$\times 10^{51}$  \\
GRB141221338 & 1.452   & 23.808  & (675.0$\pm$ 9.7)$\times 10^{-8}$ & 0.61& (353.0$\pm$ 5.1)$\times 10^{51}$ \\
GRB141221897 & 1.452   & 32.513  & (66.3$\pm$ 4.0)$\times 10^{-8}$  & 0.43 & (35.4$\pm$ 2.2)$\times 10^{51}$  \\
GRB150101641$^{\ast}$ & 0.134 & 0.08  & (222.0$\pm$ 5.8)$\times 10^{-8}$ & 0.50 & (121.0$\pm$ 3.2)$\times 10^{51}$ \\
GRB150301818 & 1.5169  & 13.312  & (244.0$\pm$ 9.0)$\times 10^{-8}$ & 0.93& (160.0$\pm$ 5.9)$\times 10^{51}$ \\
GRB150314205 & 1.758   & 10.688  & (163.0$\pm$ 8.5)$\times 10^{-8}$ & 0.96& (110.0$\pm$ 5.8)$\times 10^{51}$ \\
GRB150403913 & 2.06    & 22.272  & (72.0$\pm$ 9.7)$\times 10^{-8}$  & 1.02  & (51.2$\pm$ 6.9)$\times 10^{51}$  \\
GRB150514774 & 0.807   & 10.813  & (25.3$\pm$ 1.2)$\times 10^{-7}$  & 1.04  & (181.0$\pm$ 8.5)$\times 10^{51}$ \\
GRB150727793 & 0.313   & 49.409  & (3.22$\pm$ 1.1)$\times 10^{-7}$  & 1.03 & (24.3$\pm$ 8.5)$\times 10^{51}$  \\
GRB150821406 & 0.755   & 103.426 & (20.4$\pm$ 3.7)$\times 10^{-8}$  & 1.07& (18.2$\pm$ 3.3)$\times 10^{51}$  \\
GRB151027166$^{\$}$ & 4.063   & 123.394 & (47.0$\pm$ 4.0)$\times 10^{-8}$  & 0.69 & (43.2$\pm$ 3.6)$\times 10^{51}$  \\
GRB151111356 & 3.5     & 46.336  & (23.4$\pm$ 3.2)$\times 10^{-8}$  & 0.46 & (23.6$\pm$ 3.2)$\times 10^{51}$  \\
GRB160227831$^{\$}$ & 2.38    & 7.7    & (116.0$\pm$ 8.2)$\times 10^{-8}$ & 0.99& (120.0$\pm$ 8.5)$\times 10^{51}$ \\
GRB160303971 & 2.3     & 27.136  & (48.9$\pm$ 4.7)$\times 10^{-8}$  & 0.66& (52.5$\pm$ 5.1)$\times 10^{51}$  \\
GRB160509374 & 1.17    & 369.67  & (15.5$\pm$ 1.8)$\times 10^{-8}$  & 0.32 & (16.8$\pm$ 2.0)$\times 10^{51}$  \\
GRB160624477$^{\$}$ & 0.483   & 0.384   & (10.1$\pm$ 2.2)$\times 10^{-8}$  & 1.03 & (12.1$\pm$ 2.6 )$\times 10^{51}$  \\
GRB160625230 & 1.406   & 43.265  & (219.0$\pm$ 6.5)$\times 10^{-8}$ & 0.65& (263.0$\pm$ 7.9)$\times 10^{51}$ \\
GRB160625945 & 1.406   & 453.385 & (27.9$\pm$ 1.1)$\times 10^{-7}$  & 0.85& (35.0$\pm$ 1.3)$\times 10^{52}$  \\
GRB160821937$^{\ast}$ & 0.16    & 1.088   & (70.5$\pm$ 9.7)$\times 10^{-8}$  & 1.02 & (10.0$\pm$ 1.4)$\times 10^{52}$  \\
GRB161014522 & 2.823   & 36.609  & (3.8$\pm$ 1.2)$\times 10^{-7}$  & 0.93& (5.7$\pm$ 1.8)$\times 10^{52}$  \\
GRB161017745$^{\$}$ & 2.013   & 37.888  & (63.6$\pm$ 8.7)$\times 10^{-8}$  & 0.85 & (10.0$\pm$ 1.4)$\times 10^{52}$  \\
GRB161129300 & 0.645   & 36.096  & (31.0$\pm$ 4.0)$\times 10^{-8}$  & 1.03 & (52.2$\pm$ 6.7)$\times 10^{51}$  \\
GRB170113420 & 1.968   & 49.152  & (111.0$\pm$ 5.2)$\times 10^{-8}$ & 0.65& (190.0$\pm$ 8.8)$\times 10^{51}$ \\
GRB170214649 & 2.53    & 122.882 & (25.8$\pm$ 3.3)$\times 10^{-8}$  & 0.46 & (45.2$\pm$ 5.8)$\times 10^{51}$  \\
GRB170405777 & 3.51    & 78.593  & (6.3$\pm$ 1.1)$\times 10^{-7}$  & 1.02  & (11.1$\pm$ 1.9)$\times 10^{52}$  \\
GRB170705200$^{\$}$ & 2.01 & 27.648  & (29.9$\pm$ 3.4)$\times 10^{-8}$  & 0.61& (54.6$\pm$ 6.3)$\times 10^{51}$  \\
GRB170714049 & 0.793   & 0.224   & (242.0$\pm$ 9.6)$\times 10^{-8}$ & 0.69 & (48.6$\pm$ 1.9)$\times 10^{52}$  \\
GRB170817529 & 0.00978 & 2.048   & (24.4$\pm$ 3.0)$\times 10^{-7}$  & 0.42 & (58.8$\pm$ 7.3)$\times 10^{52}$  \\
GRB171020813 & 1.87    & 19.2    & (13.1$\pm$ 2.7)$\times 10^{-7}$& 0.59 & (31.6$\pm$ 6.5)$\times 10^{52}$  \\
GRB180620660$^{\$}$ & 1.1175  & 46.721  & (22.2$\pm$ 4.0)$\times 10^{-8}$  & 0.61& (7.1$\pm$ 1.3)$\times 10^{52}$  \\
GRB180703876 & 0.6678  & 20.736  & (15.9$\pm$ 3.3)$\times 10^{-8}$  & 1.08 & (14.3$\pm$ 2.9)$\times 10^{52}$  \\
    \hline
\end{longtable}
	\footnotesize
	{Note: The $\ast$ represent the GRBs of KN. The $\$$ represent the GRBs of EE.          
	The $\star$ represent the GRBs of SN. 
}

\bsp	
\label{lastpage}

\begin{thebibliography}{}
\makeatletter
\relax
\def\mn@urlcharsother{\let\do\@makeother \do\$\do\&\do\#\do\^\do\_\do\%\do\~}
\def\mn@doi{\begingroup\mn@urlcharsother \@ifnextchar [ {\mn@doi@}
  {\mn@doi@[]}}
\def\mn@doi@[#1]#2{\def\@tempa{#1}\ifx\@tempa\@empty \href
  {http://dx.doi.org/#2} {doi:#2}\else \href {http://dx.doi.org/#2} {#1}\fi
  \endgroup}
\def\mn@eprint#1#2{\mn@eprint@#1:#2::\@nil}
\def\mn@eprint@arXiv#1{\href {http://arxiv.org/abs/#1} {{\tt arXiv:#1}}}
\def\mn@eprint@dblp#1{\href {http://dblp.uni-trier.de/rec/bibtex/#1.xml}
  {dblp:#1}}
\def\mn@eprint@#1:#2:#3:#4\@nil{\def\@tempa {#1}\def\@tempb {#2}\def\@tempc
  {#3}\ifx \@tempc \@empty \let \@tempc \@tempb \let \@tempb \@tempa \fi \ifx
  \@tempb \@empty \def\@tempb {arXiv}\fi \@ifundefined
  {mn@eprint@\@tempb}{\@tempb:\@tempc}{\expandafter \expandafter \csname
  mn@eprint@\@tempb\endcsname \expandafter{\@tempc}}}

\bibitem[Abbott et al.(2017)]{2017ApJ...848L..13A} Abbott, B.~P., Abbott, R., Abbott, T.~D., et al.\ 2017, \apjl, 848, L13. doi:10.3847/2041-8213/aa920c
\bibitem[Aghaei Abchouyeh et al.(2023)]{2023ApJ...952..157A} Aghaei Abchouyeh, M., van Putten, M.~H.~P.~M., \& Amati, L.\ 2023, \apj, 952, 157. doi:10.3847/1538-4357/acd114
\bibitem[Becerra et al.(2015)]{2015ApJ...812..100B} Becerra, L., Cipolletta, F., Fryer, C.~L., et al.\ 2015, \apj, 812, 100. doi:10.1088/0004-637X/812/2/100
\bibitem[Bouwens et al.(2011)]{2011Natur.469..504B} Bouwens, R.~J., Illingworth, G.~D., Labbe, I., et al.\ 2011, \nat, 469, 504. doi:10.1038/nature09717
\bibitem[Butler et al.(2010)]{2010ApJ...711..495B} Butler, N.~R., Bloom, J.~S., \& Poznanski, D.\ 2010, \apj, 711, 495. doi:10.1088/0004-637X/711/1/495
\bibitem[Cao et al.(2023)]{2023SciA....9J2778C} Cao, Z., Aharonian, F., An, Q., et al.\ 2023, Science Advances, 9, eadj2778. doi:10.1126/sciadv.adj2778
\bibitem[Chen et al.(2022)]{2022ApJ...932L...7C} Chen, M.-H., Hu, R.-C., \& Liang, E.-W.\ 2022, \apjl, 932, L7. doi:10.3847/2041-8213/ac7470
\bibitem[Coward(2007)]{2007NewAR..51..539C} Coward, D.\ 2007, \nar, 51, 539. doi:10.1016/j.newar.2007.03.003
\bibitem[Dainotti et al.(2015)]{2015ApJ...800...31D} Dainotti, M.~G., Del Vecchio, R., Shigehiro, N., et al.\ 2015, \apj, 800, 31. doi:10.1088/0004-637X/800/1/31
\bibitem[Dainotti et al.(2024)]{2024ApJ...967L..30D} Dainotti, M.~G., Narendra, A., Pollo, A., et al.\ 2024, \apjl, 967, L30. doi:10.3847/2041-8213/ad4970
\bibitem[Dainotti et al.(2024)]{2024arXiv240103589D} Dainotti, M.~G., Taira, E., Wang, E., et al.\ 2024, arXiv:2401.03589. doi:10.48550/arXiv.2401.03589
\bibitem[Deng et al.(2019)]{2019JHEAp..23....1D} Deng, C.-M., Wei, J.-J., \& Wu, X.-F.\ 2019, Journal of High Energy Astrophysics, 23, 1. doi:10.1016/j.jheap.2019.05.001
\bibitem[Deng et al.(2022)]{2022ApJ...940....5D} Deng, Q., Zhang, Z.-B., Li, X.-J., et al.\ 2022, \apj, 940, 5. doi:10.3847/1538-4357/ac9590
\bibitem[Dong et al.(2022)]{2022MNRAS.513.1078D} Dong, X.~F., Li, X.~J., Zhang, Z.~B., et al.\ 2022, \mnras, 513, 1078. doi:10.1093/mnras/stac949
\bibitem[Dong et al.(2023)]{2023ApJ...958...37D} Dong, X.~F., Zhang, Z.~B., Li, Q.~M., et al.\ 2023, \apj, 958, 37. doi:10.3847/1538-4357/acf852
\bibitem[Du et al.(2025)]{2025ApJ...987L..13D} Du, X.~Y., Zhang, Z.~B., Du, W.~C., et al.\ 2025, \apjl, 987, 1, L13. doi:10.3847/2041-8213/ade1d0
\bibitem[Efron \& Petrosian(1992)]{1992ApJ...399..345E} Efron, B. \& Petrosian, V.\ 1992, \apj, 399, 345. doi:10.1086/171931
\bibitem[Gruber et al.(2014)]{2014ApJS..211...12G} Gruber, D., Goldstein, A., Weller von Ahlefeld, V., et al.\ 2014, \apjs, 211, 12. doi:10.1088/0067-0049/211/1/12
\bibitem[Hao \& Yuan(2013)]{2013A&A...558A..22H} Hao, J.-M. \& Yuan, Y.-F.\ 2013, \aap, 558, A22. doi:10.1051/0004-6361/201321471
\bibitem[Hasan \& Azzam(2024)]{2024IJAA...14...20H} Hasan, A.~M. \& Azzam, W.~J.\ 2024, International Journal of Astronomy and Astrophysics, 14, 1, 20. doi:10.4236/ijaa.2024.141002
\bibitem[Hopkins(2004)]{2004ApJ...615..209H} Hopkins, A.~M.\ 2004, \apj, 615, 209. doi:10.1086/424032
\bibitem[Hopkins \& Beacom(2006)]{2006ApJ...651..142H} Hopkins, A.~M. \& Beacom, J.~F.\ 2006, \apj, 651, 142. doi:10.1086/506610
\bibitem[Hurtado et al.(2024)]{2024ApJ...967L...4H} Hurtado, V.~U., Lloyd-Ronning, N.~M., \& Miller, J.~M.\ 2024, \apjl, 967, L4. doi:10.3847/2041-8213/ad3dfd
\bibitem[Jakobsson et al.(2006)]{2006A&A...460L..13J} Jakobsson, P., Fynbo, J.~P.~U., Ledoux, C., et al.\ 2006, \aap, 460, L13. doi:10.1051/0004-6361:20066405
\bibitem[Kang et al.(2024)]{2024MNRAS.528.5309K} Kang, Y., Liu, C., Zhu, J.-P., et al.\ 2024, \mnras, 528, 5309. doi:10.1093/mnras/stae340
\bibitem[Kouveliotou et al.(1993)]{1993ApJ...413L.101K} Kouveliotou, C., Meegan, C.~A., Fishman, G.~J., et al.\ 1993, \apjl, 413, L101. doi:10.1086/186969
\bibitem[Kyutoku et al.(2013)]{2013PhRvD..88d1503K} Kyutoku, K., Ioka, K., \& Shibata, M.\ 2013, \prd, 88, 041503. doi:10.1103/PhysRevD.88.041503
\bibitem[Lamb \& Reichart(2000)]{2000AIPC..526..658L} Lamb, D.~Q. \& Reichart, D.~E.\ 2000, Gamma-ray Bursts, 5th Huntsville Symposium, 526, 658. doi:10.1063/1.1361618
\bibitem[Lan et al.(2023)]{2023ApJ...949L...4L} Lan, L., Gao, H., Li, A., et al.\ 2023, \apjl, 949, 1, L4. doi:10.3847/2041-8213/accf93
\bibitem[Le \& Mehta(2017)]{2017ApJ...837...17L} Le, T. \& Mehta, V.\ 2017, \apj, 837, 1, 17. doi:10.3847/1538-4357/aa5fa7
\bibitem[Le et al.(2020)]{2020MNRAS.493.1479L} Le, T., Ratke, C., \& Mehta, V.\ 2020, \mnras, 493, 1, 1479. doi:10.1093/mnras/staa366
\bibitem[Lesage et al.(2023)]{2023ApJ...952L..42L} Lesage, S., Veres, P., Briggs, M.~S., et al.\ 2023, \apjl, 952, L42. doi:10.3847/2041-8213/ace5b4
\bibitem[Levan et al.(2023)]{2023NatAs...7..976L} Levan, A.~J., Malesani, D.~B., Gompertz, B.~P., et al.\ 2023, Nature Astronomy, 7, 976. doi:10.1038/s41550-023-01998-8
\bibitem[Levan et al.(2024)]{2024Natur.626..737L} Levan, A.~J., Gompertz, B.~P., Salafia, O.~S., et al.\ 2024, \nat, 626, 737. doi:10.1038/s41586-023-06759-1
\bibitem[LHAASO Collaboration et al.(2023)]{2023Sci...380.1390L} LHAASO Collaboration, Cao, Z., Aharonian, F., et al.\ 2023, Science, 380, 1390. doi:10.1126/science.adg9328
\bibitem[Li(2008)]{2008MNRAS.388.1487L} Li, L.-X.\ 2008, \mnras, 388, 1487. doi:10.1111/j.1365-2966.2008.13488.x
\bibitem[Li et al.(2024)]{2024MNRAS.527.7111L} Li, Q.~M., Sun, Q.~B., Zhang, Z.~B., et al.\ 2024, \mnras, 527, 7111. doi:10.1093/mnras/stad3619
\bibitem[Li et al.(2024)]{2024ApJ...961..201L} Li, S.-Z., Yu, Y.-W., Gao, H., et al.\ 2024, \apj, 961, 201. doi:10.3847/1538-4357/ad1593
\bibitem[Li et al.(2023)]{2023ApJ...955...98L} Li, Y., Shen, R.-F., \& Zhang, B.-B.\ 2023, \apj, 955, 98. doi:10.3847/1538-4357/acefbf
\bibitem[Liu et al.(2021)]{2021RAA....21..254L} Liu, Z.-Y., Zhang, F.-W., \& Zhu, S.-Y.\ 2021, Research in Astronomy and Astrophysics, 21, 254. doi:10.1088/1674-4527/21/10/254
\bibitem[Luo et al.(2022)]{2022MNRAS.516.1654L} Luo, J.-W., Li, Y., Ai, S., et al.\ 2022, \mnras, 516, 1654. doi:10.1093/mnras/stac2279
\bibitem[Lynden-Bell(1971)]{1971MNRAS.155...95L} Lynden-Bell, D.\ 1971, \mnras, 155, 95. doi:10.1093/mnras/155.1.95
\bibitem[Lloyd-Ronning et al.(2019)]{2019MNRAS.488.5823L} Lloyd-Ronning, N.~M., Aykutalp, A., \& Johnson, J.~L.\ 2019, \mnras, 488, 4, 5823. doi:10.1093/mnras/stz2155
\bibitem[Lloyd-Ronning et al.(2020)]{2020MNRAS.498.5041L} Lloyd-Ronning, N.~M., Johnson, J.~L., \& Aykutalp, A.\ 2020, \mnras, 498, 4, 5041. doi:10.1093/mnras/staa2787
\bibitem[Lloyd-Ronning et al.(2024)]{2024MNRAS.535.2800L} Lloyd-Ronning, N.~M., Johnson, J., Upton Sanderbeck, P., et al.\ 2024, \mnras, 535, 3, 2800. doi:10.1093/mnras/stae2502
\bibitem[Narayan et al.(1992)]{1992ApJ...395L..83N} Narayan, R., Paczynski, B., \& Piran, T.\ 1992, \apjl, 395, L83. doi:10.1086/186493
\bibitem[Paczy{\'n}ski(1998)]{1998ApJ...494L..45P} Paczy{\'n}ski, B.\ 1998, \apjl, 494, L45. doi:10.1086/311148
\bibitem[Patricelli \& Bernardini(2020)]{2020MNRAS.499L..96P} Patricelli, B. \& Bernardini, M.~G.\ 2020, \mnras, 499, L96. doi:10.1093/mnrasl/slaa169
\bibitem[Paul(2018)]{2018MNRAS.477.4275P} Paul, D.\ 2018, \mnras, 477, 4275. doi:10.1093/mnras/sty840
\bibitem[Pescalli et al.(2016)]{2016A&A...587A..40P} Pescalli, A., Ghirlanda, G., Salvaterra, R., et al.\ 2016, \aap, 587, A40. doi:10.1051/0004-6361/201526760
\bibitem[Perley et al.(2016)]{2016ApJ...817....7P} Perley, D.~A., Kr{\"u}hler, T., Schulze, S., et al.\ 2016, \apj, 817, 1, 7. doi:10.3847/0004-637X/817/1/7
\bibitem[Petrosian et al.(2015)]{2015ApJ...806...44P} Petrosian, V., Kitanidis, E., \& Kocevski, D.\ 2015, \apj, 806, 1, 44. doi:10.1088/0004-637X/806/1/44
\bibitem[Petrosian \& Dainotti(2024)]{2024ApJ...963L..12P} Petrosian, V. \& Dainotti, M.~G.\ 2024, \apjl, 963, 1, L12. doi:10.3847/2041-8213/ad2763
\bibitem[Porciani \& Madau(2001)]{2001ApJ...548..522P} Porciani, C. \& Madau, P.\ 2001, \apj, 548, 522. doi:10.1086/319027
\bibitem[Rahin \& Behar(2019)]{2019ApJ...885...47R} Rahin, R. \& Behar, E.\ 2019, \apj, 885, 47. doi:10.3847/1538-4357/ab3e34
\bibitem[Rossi et al.(2022)]{2022ApJ...932....1R} Rossi, A., Rothberg, B., Palazzi, E., et al.\ 2022, \apj, 932, 1. doi:10.3847/1538-4357/ac60a2
\bibitem[Rueda \& Ruffini(2012)]{2012ApJ...758L...7R} Rueda, J.~A. \& Ruffini, R.\ 2012, \apjl, 758, L7. doi:10.1088/2041-8205/758/1/L7
\bibitem[Ruffini(2013)]{2013IJMPD..2260009R} Ruffini, R.\ 2013, International Journal of Modern Physics D, 22, 1360009. doi:10.1142/S0218271813600092
\bibitem[Salvaterra et al.(2012)]{2012ApJ...749...68S} Salvaterra, R., Campana, S., Vergani, S.~D., et al.\ 2012, \apj, 749, 68. doi:10.1088/0004-637X/749/1/68
\bibitem[Singal et al.(2011)]{2011ApJ...743..104S} Singal, J., Petrosian, V., Lawrence, A., et al.\ 2011, \apj, 743, 104. doi:10.1088/0004-637X/743/2/104
\bibitem[Sun et al.(2015)]{2015ApJ...812...33S} Sun, H., Zhang, B., \& Li, Z.\ 2015, \apj, 812, 33. doi:10.1088/0004-637X/812/1/33
\bibitem[Thompson et al.(2006)]{2006ApJ...647..787T} Thompson, R.~I., Eisenstein, D., Fan, X., et al.\ 2006, \apj, 647, 787. doi:10.1086/505568
\bibitem[Totani(1997)]{1997ApJ...486L..71T} Totani, T.\ 1997, \apjl, 486, L71. doi:10.1086/310853
\bibitem[Tsvetkova et al.(2021)]{2021ApJ...908...83T} Tsvetkova, A., Frederiks, D., Svinkin, D., et al.\ 2021, \apj, 908, 1, 83. doi:10.3847/1538-4357/abd569
\bibitem[van Putten \& Della Valle(2023)]{2023A&A...669A..36V} van Putten, M.~H.~P.~M. \& Della Valle, M.\ 2023, \aap, 669, A36. doi:10.1051/0004-6361/202142974
\bibitem[Wang et al.(2020)]{2020ApJ...902L..42W} Wang, J.-S., Peng, Z.-K., Zou, J.-H., et al.\ 2020, \apjl, 902, L42. doi:10.3847/2041-8213/abbfb8
\bibitem[Wijers et al.(1998)]{1998MNRAS.294L..13W} Wijers, R.~A.~M.~J., Bloom, J.~S., Bagla, J.~S., et al.\ 1998, \mnras, 294, L13. doi:10.1046/j.1365-8711.1998.01328.x10.1111/j.1365-8711.1998.01328.x
\bibitem[Woosley(1993)]{1993ApJ...405..273W} Woosley, S.~E.\ 1993, \apj, 405, 273. doi:10.1086/172359
\bibitem[Woosley \& Bloom(2006)]{2006ARA&A..44..507W} Woosley, S.~E. \& Bloom, J.~S.\ 2006, \araa, 44, 507. doi:10.1146/annurev.astro.43.072103.150558
\bibitem[Yin et al.(2023)]{2023ApJ...954L..17Y} Yin, Y.-H.~I., Zhang, B.-B., Sun, H., et al.\ 2023, \apjl, 954, L17. doi:10.3847/2041-8213/acf04a
\bibitem[Yu et al.(2015)]{2015ApJS..218...13Y} Yu, H., Wang, F.~Y., Dai, Z.~G., et al.\ 2015, \apjs, 218, 13. doi:10.1088/0067-0049/218/1/13
\bibitem[Zeng et al.(2021)]{2021ApJ...913..120Z} Zeng, H., Petrosian, V., \& Yi, T.\ 2021, \apj, 913, 120. doi:10.3847/1538-4357/abf65e
\bibitem[Zhang et al.(2021)]{2021NatAs...5..911Z} Zhang, B.-B., Liu, Z.-K., Peng, Z.-K., et al.\ 2021, Nature Astronomy, 5, 911. doi:10.1038/s41550-021-01395-z
\bibitem[Zhang \& M{\'e}sz{\'a}ros(2004)]{2004IJMPA..19.2385Z} Zhang, B. \& M{\'e}sz{\'a}ros, P.\ 2004, International Journal of Modern Physics A, 19, 2385. doi:10.1142/S0217751X0401746X
\bibitem[Zhang et al.(2025)]{2025JHEAp..45..392Z} Zhang, B.~T., Murase, K., Ioka, K., et al.\ 2025, Journal of High Energy Astrophysics, 45, 392. doi:10.1016/j.jheap.2025.01.007
\bibitem[Zhang \& Wang(2018)]{2018ApJ...852....1Z} Zhang, G.~Q. \& Wang, F.~Y.\ 2018, \apj, 852, 1. doi:10.3847/1538-4357/aa9ce5
\bibitem[Zhang et al.(2025)]{2025A&A...698A..18Z} Zhang, K.~J., Zhang, Z.~B., Rodin, A.~E., et al.\ 2025, \aap, 698, A18. doi:10.1051/0004-6361/202450727
\bibitem[Zhang et al.(2021)]{2021MNRAS.501..157Z} Zhang, R.~C., Zhang, B., Li, Y., et al.\ 2021, \mnras, 501, 1, 157. doi:10.1093/mnras/staa3537
\bibitem[Zhang \& Choi(2008)]{2008A&A...484..293Z} Zhang, Z.-B. \& Choi, C.-S.\ 2008, \aap, 484, 293. doi:10.1051/0004-6361:20079210
\bibitem[Zhang et al.(2020)]{2020ApJ...902...40Z} Zhang, Z.~B., Jiang, M., Zhang, Y., et al.\ 2020, \apj, 902, 40. doi:10.3847/1538-4357/abb400
\bibitem[Zhang et al.(2018)]{2018PASP..130e4202Z} Zhang, Z.~B., Zhang, C.~T., Zhao, Y.~X., et al.\ 2018, \pasp, 130, 054202. doi:10.1088/1538-3873/aaa6af
\bibitem[Zhu et al.(2021)]{2021ApJ...917...24Z} Zhu, J.-P., Wu, S., Yang, Y.-P., et al.\ 2021, \apj, 917, 24. doi:10.3847/1538-4357/abfe5e

\makeatother
\end{thebibliography}
\end{document}